\PassOptionsToPackage{unicode}{hyperref}
\PassOptionsToPackage{hyphens}{url}
\PassOptionsToPackage{dvipsnames,svgnames,x11names}{xcolor}
\documentclass[
  12pt]{article}

\usepackage{amsmath,amssymb}
\usepackage{iftex}
\ifPDFTeX
  \usepackage[T1]{fontenc}
  \usepackage[utf8]{inputenc}
  \usepackage{textcomp} % provide euro and other symbols
\else % if luatex or xetex
  \usepackage{unicode-math}
  \defaultfontfeatures{Scale=MatchLowercase}
  \defaultfontfeatures[\rmfamily]{Ligatures=TeX,Scale=1}
\fi
\usepackage{lmodern}
\ifPDFTeX\else  
\fi
\IfFileExists{upquote.sty}{\usepackage{upquote}}{}
\IfFileExists{microtype.sty}{% use microtype if available
  \usepackage[]{microtype}
  \UseMicrotypeSet[protrusion]{basicmath} % disable protrusion for tt fonts
}{}
\makeatletter
\@ifundefined{KOMAClassName}{% if non-KOMA class
  \IfFileExists{parskip.sty}{%
    \usepackage{parskip}
  }{% else
    \setlength{\parindent}{0pt}
    \setlength{\parskip}{6pt plus 2pt minus 1pt}}
}{% if KOMA class
  \KOMAoptions{parskip=half}}
\makeatother
\usepackage{xcolor}
\makeatletter
\ifx\paragraph\undefined\else
  \let\oldparagraph\paragraph
  \renewcommand{\paragraph}{
    \@ifstar
      \xxxParagraphStar
      \xxxParagraphNoStar
  }
  \newcommand{\xxxParagraphStar}[1]{\oldparagraph*{#1}\mbox{}}
  \newcommand{\xxxParagraphNoStar}[1]{\oldparagraph{#1}\mbox{}}
\fi
\ifx\subparagraph\undefined\else
  \let\oldsubparagraph\subparagraph
  \renewcommand{\subparagraph}{
    \@ifstar
      \xxxSubParagraphStar
      \xxxSubParagraphNoStar
  }
  \newcommand{\xxxSubParagraphStar}[1]{\oldsubparagraph*{#1}\mbox{}}
  \newcommand{\xxxSubParagraphNoStar}[1]{\oldsubparagraph{#1}\mbox{}}
\fi
\makeatother

\usepackage{longtable,booktabs,array}
\usepackage{calc} % for calculating minipage widths
\usepackage{etoolbox}
\makeatletter
\patchcmd\longtable{\par}{\if@noskipsec\mbox{}\fi\par}{}{}
\makeatother
\IfFileExists{footnotehyper.sty}{\usepackage{footnotehyper}}{\usepackage{footnote}}
\makesavenoteenv{longtable}
\usepackage{graphicx}
\makeatletter
\def\maxwidth{\ifdim\Gin@nat@width>\linewidth\linewidth\else\Gin@nat@width\fi}
\def\maxheight{\ifdim\Gin@nat@height>\textheight\textheight\else\Gin@nat@height\fi}
\makeatother
\setkeys{Gin}{width=\maxwidth,height=\maxheight,keepaspectratio}
\makeatletter
\def\fps@figure{htbp}
\makeatother

\makeatletter
\@ifpackageloaded{caption}{}{\usepackage{caption}}
\AtBeginDocument{%
\ifdefined\contentsname
  \renewcommand*\contentsname{Table of contents}
\else
  \newcommand\contentsname{Table of contents}
\fi
\ifdefined\listfigurename
  \renewcommand*\listfigurename{List of Figures}
\else
  \newcommand\listfigurename{List of Figures}
\fi
\ifdefined\listtablename
  \renewcommand*\listtablename{List of Tables}
\else
  \newcommand\listtablename{List of Tables}
\fi
\ifdefined\figurename
  \renewcommand*\figurename{Figure}
\else
  \newcommand\figurename{Figure}
\fi
\ifdefined\tablename
  \renewcommand*\tablename{Table}
\else
  \newcommand\tablename{Table}
\fi
}
\@ifpackageloaded{float}{}{\usepackage{float}}
\floatstyle{ruled}
\@ifundefined{c@chapter}{\newfloat{codelisting}{h}{lop}}{\newfloat{codelisting}{h}{lop}[chapter]}
\floatname{codelisting}{Listing}

\makeatother
\makeatletter
\@ifpackageloaded{caption}{}{\usepackage{caption}}
\@ifpackageloaded{subcaption}{}{\usepackage{subcaption}}
\makeatother

\ifLuaTeX
  \usepackage{selnolig}  % disable illegal ligatures
\fi
\usepackage[]{natbib}
\usepackage{bookmark}

\IfFileExists{xurl.sty}{\usepackage{xurl}}{} % add URL line breaks if available
\hypersetup{
  pdftitle={Title},
  pdfauthor={Author 1; Author 2},
  pdfkeywords={3 to 6 keywords, that do not appear in the title},
  colorlinks=true,
  linkcolor={blue},
  filecolor={Maroon},
  citecolor={Blue},
  urlcolor={Blue},
  pdfcreator={LaTeX via pandoc}}

\newcommand{\anon}{1}

\begin{document}

\def\spacingset#1{\renewcommand{\baselinestretch}%
{#1}\small\normalsize} \spacingset{1}

%%%%%%%%%%%%%%%%%%%%%%%%%%%%%%%%%%%%%%%%%%%%%%%%%%%%%%%%%%%%%%%%%%%%%%%%%%%%%%

\if1\anon
{
  % \title{\bf Online Reliability Prediction for Satellite Electronics via Spatiotemporal Active Learning}
  \title{\bf Adaptive Active Learning for Online Reliability Prediction of Satellite Electronics}
  \author{Shixiang Li\textsuperscript{1}\\
     ZJUI Institute, Zhejiang University\\
     Yubin Tian\textsuperscript{1}\\
     Faculty of Computational Mathematics and Cybernetics, \\ Shenzhen MSU-BIT University\\
     Dianpeng Wang\textsuperscript{2}\thanks{wdp@bit.edu.cn}\hspace{.2cm} \\
    The School of Mathematics and Statistics, Beijing Institute of Technology \\
    Piao Chen\textsuperscript{3}\\
     ZJUI Institute, Zhejiang University\\
    and \\
    Mengying Ren\textsuperscript{4}\\
     The School of Mathematics and Statistics, Beijing Institute of Technology
    }
  \maketitle
} \fi

\if0\anon
{
  \bigskip
  \bigskip
  \bigskip
  \begin{center}
    %{\LARGE\bf Online Reliability Prediction for Satellite Electronics via Spatiotemporal Active Learning}
    {\LARGE\bf Adaptive Active Learning for Online Reliability Prediction of Satellite Electronics}
\end{center}
  \medskip
} \fi

\bigskip
\begin{abstract}
Accurate on-orbit reliability prediction for satellite electronics is often hindered by limited data availability, varying operational conditions, and considerable unit-to-unit variability. 
To overcome these obstacles, this paper proposes a novel integrated online reliability prediction framework. 
The main contributions are twofold. 
First, a Wiener process-based degradation model is developed, incorporating a generalized Arrhenius link function, individual random effects, and spatial correlations among adjacent units. 
A customized maximum likelihood estimation method is further devised to facilitate efficient and accurate parameter inference. 
Second, a two-stage active learning sampling scheme is designed to adaptively enhance prediction accuracy. 
This strategy initially selects representative units based on spatial configuration, and subsequently determines optimal sampling times using a comprehensive criterion that balances unit-specific information, model uncertainty, and degradation dynamics. 
Numerical experiments and a practical case study from the Tiangong space station demonstrate that the proposed method markedly improves reliability prediction accuracy while significantly reducing data requirements, offering an efficient solution for the prognostic and health management of complex satellite electronic systems.
\end{abstract}

\noindent%
{\it Keywords:} D-optimal, lifetime, optimized sampling, reliability analysis, Wiener process
\vfill

\newpage
\spacingset{1.8} % DON'T change the spacing!

\section{Introduction}
Reliability assessment is indispensable for ensuring the safety and function of complex engineering systems \citep{wang2025reliability}. 
Modern systems are increasingly designed with high reliability and extended service lifespans to maximize mission value. 
Therefore, corresponding failure events are extremely rare within feasible observation windows \citep{ye2015stochastic}, posing a fundamental challenge to traditional reliability analysis based on failure data and lifetime models \citep{zhang2022wiener}.
Meanwhile, with the rapid development of online monitoring technologies, various sensors can be used to record the health status \citep{wang2015preventive}.
Thus, degradation modeling has been studied as an alternative, which leverages continuous performance data to infer the true current degradation state \citep{zhu2026online}.
In the degradation analysis framework, the health status of a unit is characterized by a measurable performance characteristic $X(t)$ that degrades over time \citep{zhai2025modeling}. 
A unit is considered to experience a failure when $X(t)$ first crosses a predefined critical threshold \citep{Li29082025}. 
By utilizing degradation data, it is possible to predict the reliability, promoting preventive maintenance, repair, or replacement to avoid the transition from potential defects to actual failures \citep{si2013wiener,yang2018hybrid,wang2023degradation}.

Existing literature broadly classifies degradation modeling into general path approaches and stochastic process models. 
General path models utilize regression frameworks under known physical knowledge and have wide applications, including \citep{lu2021general,wang2024flexible,ye2024path}.
On the other hand, stochastic process models, using independent increments, naturally capture the temporal variation \citep{kang2023reliability}. 
The Wiener process, in particular, stands out as a popular choice among alternatives like Gamma \citep{chen2018uncertainty} or Inverse Gaussian processes \citep{xu2025efficient}. 
Its suitability is based on the Central Limit Theorem, which suggests that degradation increments, representing the cumulative sum of numerous micro-level influences, tend to be normally distributed \citep{wang2013stochastic, zhang2022wiener}.
However, while such methods have been widely applied \citep{zhang2018degradation,fang2024class,chen2025condition}, their direct implementation on long-endurance aerospace assets poses different challenges, as exemplified by the China Tiangong Space Station (CTSS).

The space station, a strategic asset designed for long-term on-orbit research \citep{wang2023design, yin2023orbit}, exemplifies the complexity of reliability analysis.
This study targets the core switching components within the space station's power distribution units (PDUs), specifically the Metal-Oxide-Semiconductor Field-Effect Transistors (MOSFETs), which are responsible for dynamic unit regulation and voltage conversion \citep{zarghany2016fault} under high-stakes operational constraints.
As noted by \citep{xu2023reliability} and \citep{zeng2023reliability}, such units are 
crucial subsystems in satellites.
Accurately predicting the reliability of these critical components throughout their extended lifecycles presents several unique challenges that necessitate advanced modeling frameworks.

However, a critical review of the literature reveals two significant gaps that motivate this work.
The first gap lies in degradation modeling under the unique conditions of a space station. 
The operating environment is complex and highly dynamic. 
Unlike short-term missions, the space station experiences continuous alternating orbital cycles, which induce fluctuations in junction temperature \citep{xu2025reliability}. 
Coupled with varying electrical stresses from shifting payload operations \citep{musallam2014mission}, these factors introduce significant nonlinearity into degradation trajectories, rendering commonly used constant-stress models ineffective. 
Furthermore, the system exhibits inherent unit-wise heterogeneity due to manufacturing tolerances. 
While recent studies have begun to address such complexities by incorporating environmental covariates \citep{zhang2018degradation} and random effects to capture parameter variability \citep{pennell2010bayesian,xu2016nonlinear,duan2018exponential}, they overlook a critical aspect: the spatial dependence among units. 
Due to the compact spatial topology of the Power Distribution Unit (PDU), adjacent units exhibit strong coupling effects, where the operation of one unit inevitably influences its spatial neighbors. 
Existing models neglect this inherent spatial dependence structure, which, as will be demonstrated, can lead to misleading reliability predictions.

The second gap pertains to data acquisition. 
Strictly limited data transmission resources make high-frequency, full-scale monitoring infeasible, necessitating a focused active learning strategy to capture the most valuable data. 
While the literature on general sampling design is extensive \citep{lee2022failure, sun2023uniform, li2023parallel, tai2025d}, only a subset addresses the optimization of observation epochs for degradation prediction \citep{wang2018adaptive, wang2022adaptive, ren2025sequential}. 
Some research focuses predominantly on step-stress accelerated degradation test plans rather than observation time \citep{tseng2009optimal, tsai2012optimal, cheng2024optimal}. Critically, these time-focused approaches neglect the spatial dimension and fail to select informative units based on environmental and spatial contexts. 
In our study, simultaneous observation of all units is unnecessary, creating a dual optimization problem: identifying the most informative units according to the unit topology (spatial selection) while concurrently determining the optimal observation times (temporal scheduling).

Given the high stakes, where any unexpected failure could lead to severe economic losses or compromise living conditions, accurate reliability prediction is essential. 
To bridge these gaps, this article proposes an integrated online reliability prediction method for satellite electronics. 
The main contributions are twofold: (1) a degradation model that jointly incorporates individual heterogeneity and spatial dependence, which is an aspect rarely considered in existing literature; and (2) a new spatiotemporal active learning sampling strategy that optimizes data acquisition across both units and observation epochs, a challenge unique to the space station context.

The remainder of this paper is organized as follows. 
Section~\ref{sec:model} establishes the Wiener-based degradation model considering time-varying stresses, unit-wise heterogeneity, and spatial correlations. 
Section~\ref{sec:est} details the efficient parameter estimation method. 
Section~\ref{sec:design} presents the two-stage spatiotemporal active learning sampling strategy. 
Sections~\ref{sec:example} demonstrate the effectiveness of the proposed method through numerical simulations and a practical application involving the Tiangong space station. 
Finally, Section~\ref{sec:con} concludes the paper.
% All technique details and additional numerical results are provided in the Appendix.

\section{A Hierarchical Spatiotemporal Degradation Model}
\label{sec:model}
Satellite electronics operate under conditions that challenge conventional degradation models: environments are dynamic, units exhibit pronounced heterogeneity, and compact physical layouts induce spatial dependence among adjacent components. 
This section develops a novel degradation framework that simultaneously captures all three features. 
We first formulate the degradation process for a single unit, incorporating time‑varying covariates and random effects. 
The model is then extended to the entire system by explicitly integrating a spatial correlation structure, linking physically adjacent units through their random coefficients.

Let $X_i(t)$ denote the performance characteristic of the $i$th unit ($i = 1, 2, \dots, L$) at time $t$. 
We adopt a Wiener process with a time‑scale transformation,
\begin{equation} 
\label{eq:base_model}
X_i(t) = X_{i,0} + \beta_i(t)\Lambda(t) + \sigma \mathbb{B}_i(\Lambda(t)),
\end{equation}
where $X_{i,0}$ is the initial degradation level (set to $0$ without loss of generality), $\beta_i(t)$ is a cumulative degradation coefficient that accumulates damage under dynamic environments, $\Lambda(t) = t^\alpha$ ($\alpha > 0$) is a continuous non‑decreasing function representing the intrinsic nonlinear degradation path, $\sigma > 0$ is the diffusion coefficient, and $\mathbb{B}_i(\cdot)$ denotes standard Brownian motion, independent across units.

Given the dynamic operating environment, both junction temperature $S_{i,1}(t)$ and electrical stress $S_{i,2}(t)$ fluctuate according to known orbital and operational patterns. 
To capture their impact on degradation, we adopt an approach analogous to the proportional hazards framework, modeling the degradation rate as an exponential function of dynamic covariates, which is given by
\begin{align*}
\beta_i(t) &= a_i \exp\bigl[\gamma_1 Z_{i,1}(t) + \gamma_2 Z_{i,2}(t)\bigr],\\
Z_{i,1}(t) &= \frac{1000}{S_{i,1}(t) + 273.15}, \quad 
Z_{i,2}(t) = \ln S_{i,2}(t),
\end{align*}
where $\gamma_1$ and $\gamma_2$ are acceleration coefficients. 
The unit‑specific random coefficient $a_i \sim N(\mu_a, \tau_a^2)$ captures unobserved individual differences arising from manufacturing tolerances. 
Consequently, the degradation rate follows a time‑dependent normal distribution given by
$$
\beta_i(t) \sim N\bigl(\mu_{i,b}(t),\; \tau_{i,b}^2(t)\bigr),
$$
with mean $\mu_{i,b}(t) = \mu_a \lambda_i(t)$, variance $\tau_{i,b}^2(t) = \tau_a^2 \lambda_i^2(t)$, and $\lambda_i(t) = \exp\bigl[\gamma_1 Z_{i,1}(t) + \gamma_2 Z_{i,2}(t)\bigr]$ encapsulating the environmental effects.

From the model~\eqref{eq:base_model}, $X_i(t)$ is normally distributed with moments
\begin{align*}
\mathbb{E}[X_i(t)] &= \mu_a \lambda_i(t) \Lambda(t), \\
\mathbb{V}ar[X_i(t)] &= \tau_a^2 [\lambda_i(t) \Lambda(t)]^2 + \sigma^2 \Lambda(t), \label{eq:var}\\
\mathbb{C}ov[X_i(t), X_i(s)] &= \tau_a^2 [\lambda_i(t)\Lambda(t)][\lambda_i(s)\Lambda(s)] + \sigma^2 \Lambda(t \wedge s),
\end{align*}
where $t \wedge s = \min(t, s)$. 
These expressions reveal two distinct sources of variability: the random drift term $\tau_a^2$ captures persistent unit‑specific deviations, while the diffusion term $\sigma^2$ accounts for transient fluctuations.

A key innovation of this work is the explicit modeling of spatial dependencies. 
In the space station layout, adjacent units share similar thermal and electrical microenvironments. 
Suppose the units are arranged on a one‑dimensional spatial domain and indexed by location $i$. 
We assume that the Brownian motions are independent across units, but the inherent degradation rates are correlated. 
The covariance between the random coefficients $a_i$ and $a_j$ is modeled as
\begin{equation*}
\mathbb{C}ov(a_i, a_j) = 
\begin{cases}
\tau_a^2, & i = j,\\
\tau_a^2 \rho, & |i - j| = 1 \text{ (adjacent)},\\
0, & \text{otherwise},
\end{cases}
\end{equation*}
where $\rho \in (-1,1)$ quantifies the spatial correlation strength. This first‑order autoregressive structure is both parsimonious and physically motivated: correlation decays with distance and vanishes beyond immediate neighbors, reflecting the localized nature of thermal and electrical coupling.

Let $\boldsymbol{Y} = \bigl(X_1(t_1), X_2(t_2), \dots, X_L(t_L)\bigr)^\top$ denote the column vector of observations from all units, where $t_i$ represents the observation time for unit $i$. 
The joint distribution of $\boldsymbol{Y}$ is multivariate normal, with mean vector $\bigl(\mu_a \lambda_1(t_1)\Lambda(t_1), \dots, \mu_a \lambda_L(t_L)\Lambda(t_L)\bigr)^\top$ and covariance matrix $\boldsymbol{\Psi}$. 
Its $(i,j)$th element, representing the covariance between units $i$ and $j$, is
\begin{equation}
\label{eq:global_cov_element}
\boldsymbol{\Psi}_{i,j} = \mathbb{C}ov(a_i, a_j)\, \lambda_i(t_i)\Lambda(t_i)\, \lambda_j(t_j)\Lambda(t_j) + \delta_{ij} \sigma^2 \Lambda(t_i \wedge t_j),
\end{equation}
where $\delta_{ij}$ is the Kronecker delta ($\delta_{ij}=1$ if $i=j$, and $0$ otherwise). 
Equation~\eqref{eq:global_cov_element} decomposes system variability into three physically interpretable components: spatial correlation governed by $\rho$, temporal fluctuations governed by $\sigma^2$, and unit‑wise heterogeneity governed by $\tau_a^2$.

For reliability prediction, we focus on individual units. 
The lifetime $T_i$ of unit $i$ is defined as the first hitting time of a critical failure threshold $\xi$, namely
\begin{equation*}
T_i = \inf\{ t \ge 0 : X_i(t) \ge \xi \}.
\end{equation*}
Because $\beta_i(t)$ is time‑varying, a closed‑form expression for the distribution of $T_i$ is mathematically intractable. 
We therefore employ Monte Carlo simulation for lifetime estimation in subsequent sections.

% {\color{red}\textbf{Note that (check correct?):}
Note that the proposed framework generalizes several existing models.
For instance, setting $\rho = 0$ simplifies the framework to a collection of independent units with random effects. 
When $\tau_a^2 = 0$, it becomes a fixed‑effects model with spatial independence. 
Furthermore, under constant environmental covariates, $\lambda_i(t)$ reduces to a constant, recovering traditional Wiener process models. 
This nested structure facilitates model comparison and hypothesis testing.

\section{Efficient Profile Likelihood Inference}
\label{sec:est}
Estimating the parameters of a spatiotemporal degradation model poses considerable challenges: the likelihood involves high‑dimensional covariance matrices, and the number of parameters is large. 
In this section, we develop an efficient estimation procedure that exploits the separable structure of the model to reduce the dimensionality of numerical optimization. 
The key idea is to concentrate out the scale parameters analytically, leaving only a low‑dimensional profile likelihood to be maximized.

Consider degradation data from $L$ spatially dependent units observed over the time interval $[0, \mathcal{T}]$. 
Measurements are taken at time points $\boldsymbol{t}_i = \{t_{i,1},\dots,t_{i,m_i}\}$ for unit $i$ ($i = 1,\dots,L$), yielding observations $x_{i,j} = X_i(t_{i,j})$, $j = 1,\dots,m_i$. 
The complete dataset is denoted $\mathcal{D} = \{(t_{i,j}, x_{i,j}) : i = 1,\dots,L,\; j = 1,\dots,m_i\}$.

Let $\boldsymbol{x}_i = (x_{i,1}, \dots, x_{i,m_i})^\top$ be the observation vector for unit $i$. 
Stacking all observations gives the system‑level vector $\boldsymbol{y} = (\boldsymbol{x}_1^\top, \boldsymbol{x}_2^\top, \dots, \boldsymbol{x}_L^\top)^\top \in \mathbb{R}^M$, where $M = \sum_{i=1}^L m_i$ is the total number of measurements.
From the model, $\mathbb{E}[\boldsymbol{x}_i] = \mu_a \boldsymbol{\Xi}_i$, with $\boldsymbol{\Xi}_i^{(j)} = \lambda_i(t_{i,j})\Lambda(t_{i,j})$. 
Hence, $\mathbb{E}[\boldsymbol{y}] = \mu_a \boldsymbol{\Xi}$, where $\boldsymbol{\Xi} = (\boldsymbol{\Xi}_1^\top, \boldsymbol{\Xi}_2^\top, \dots, \boldsymbol{\Xi}_L^\top)^\top$.
According to the dependence structure derived in Section~\ref{sec:model}, the global covariance matrix $\boldsymbol{\Psi} \in \mathbb{R}^{M \times M}$ consists of $L \times L$ block matrices. 
Because Brownian motions are independent across units, the temporal variability term appears only in the diagonal blocks. 
Conversely, spatial correlation, governed by the random drift coefficients, affects both diagonal and off‑diagonal blocks. 
Thus, the $(i,j)$th block matrix $\boldsymbol{\Psi}_{i,j}$ is
\begin{equation*}
\boldsymbol{\Psi}_{i,j} = 
\begin{cases}
\tau_a^2 \bigl(\boldsymbol{\Xi}_i \boldsymbol{\Xi}_i^\top + \kappa^2 \boldsymbol{Q}_i\bigr), & i = j,\\
\tau_a^2 \rho \,\boldsymbol{\Xi}_i \boldsymbol{\Xi}_j^\top, & |i - j| = 1,\\
\boldsymbol{0}, & \text{otherwise},
\end{cases}
\end{equation*}
where $\kappa = \sigma / \tau_a$ is a reparameterized diffusion coefficient, and $\boldsymbol{Q}_i$ represents the Brownian motion covariance structure for unit $i$, with entries $\boldsymbol{Q}_i^{(l,k)} = \min\{\Lambda(t_{i,l}), \Lambda(t_{i,k})\}$.

For mathematical convenience, we factor the covariance matrix as $\boldsymbol{\Psi} = \tau_a^2 \widetilde{\boldsymbol{\Psi}}$, where the scaled matrix $\widetilde{\boldsymbol{\Psi}}$ is defined blockwise.
\begin{equation*}
\widetilde{\boldsymbol{\Psi}}_{i,j} = 
\begin{cases}
\boldsymbol{\Xi}_i \boldsymbol{\Xi}_i^\top + \kappa^2 \boldsymbol{Q}_i, & i = j,\\
\rho \,\boldsymbol{\Xi}_i \boldsymbol{\Xi}_j^\top, & |i - j| = 1,\\
\boldsymbol{0}, & \text{otherwise}.
\end{cases}
\end{equation*}

Let $\boldsymbol{\theta}=(\alpha, \mu_a, \tau_a^2, \kappa, \gamma_1, \gamma_2, \rho)^\top$ denote the complete parameter vector.
The log‑likelihood function for $\boldsymbol{\theta}$ given $\mathcal{D}$ is
\begin{equation}
\label{eq:likelihood_y}
\begin{aligned}
l(\boldsymbol{\theta} \mid \mathcal{D}) \propto  -\frac{M}{2}\ln\tau_a^2 - \frac{1}{2}\ln|\widetilde{\boldsymbol{\Psi}}| 
&- \frac{1}{2\tau_a^2}(\boldsymbol{y} - \mu_a \boldsymbol{\Xi})^\top \widetilde{\boldsymbol{\Psi}}^{-1}(\boldsymbol{y} - \mu_a \boldsymbol{\Xi}).
\end{aligned}
\end{equation}

Direct maximization over the full parameter space can be computationally unstable due to high dimensionality. 
However, the optimization problem exhibits a separable structure: for any fixed set of structural parameters $\boldsymbol{\theta}_1 = (\alpha, \kappa, \gamma_1, \gamma_2, \rho)^\top$, the maximum likelihood estimates of $\mu_a$ and $\tau_a^2$ can be derived analytically. 
Exploiting this property, we employ the profile likelihood method \citep{murphy2000profile} to concentrate out $\mu_a$ and $\tau_a^2$, thereby reducing the dimensionality of numerical optimization.

Given $\boldsymbol{\theta}_1$, the closed‑form estimates are
\begin{align}
\widehat{\mu}_a(\boldsymbol{\theta}_1) &= \frac{\boldsymbol{\Xi}^\top \widetilde{\boldsymbol{\Psi}}^{-1} \boldsymbol{y}}{\boldsymbol{\Xi}^\top \widetilde{\boldsymbol{\Psi}}^{-1} \boldsymbol{\Xi}}, \label{eq:mu_hat}\\
\widehat{\tau}_a^2(\boldsymbol{\theta}_1) &= \frac{1}{M} (\boldsymbol{y} - \widehat{\mu}_a(\boldsymbol{\theta}_1)\boldsymbol{\Xi})^\top \widetilde{\boldsymbol{\Psi}}^{-1} (\boldsymbol{y} - \widehat{\mu}_a(\boldsymbol{\theta}_1)\boldsymbol{\Xi}). \label{eq:tau_hat}
\end{align}
Substituting these into \eqref{eq:likelihood_y} yields the profile log‑likelihood function, which depends solely on $\boldsymbol{\theta}_1$.
\begin{equation}
\label{eq:profilelikelihood}
l_p(\boldsymbol{\theta}_1 \mid \boldsymbol{y}) \propto  - \frac{M}{2}\ln\widehat{\tau}_a^2(\boldsymbol{\theta}_1) - \frac{1}{2}\ln|\widetilde{\boldsymbol{\Psi}}|.
\end{equation}

Estimation of $\boldsymbol{\theta}_1$ proceeds by maximizing \eqref{eq:profilelikelihood}. 
This is a constrained nonlinear optimization problem, with $\kappa > 0$ and $\rho \in [-1,1]$. 
To ensure numerical stability and computational efficiency, especially for the log‑determinant $\ln|\widetilde{\boldsymbol{\Psi}}|$ involving large matrices, we employ the Cholesky decomposition of $\widetilde{\boldsymbol{\Psi}}$. 
The optimization is performed using numerical methods with a multi‑start strategy to mitigate the risk of converging to local optima. 
Once $\widehat{\boldsymbol{\theta}}_1$ is obtained, the estimates for $\mu_a$ and $\tau_a^2$ are recovered from \eqref{eq:mu_hat} and \eqref{eq:tau_hat}.

\section{Optimal Spatiotemporal Data Acquisition via Active Learning}
\label{sec:design}
On‑orbit satellite data are severely constrained by limited transmission bandwidth. 
Consequently, an efficient active learning sampling strategy is essential for monitoring degradation and predicting reliability. 
This section addresses two intertwined problems: spatial sampling which selects a representative subset of units to observe at each epoch and sequential temporal sampling involving determining the optimal time for the next observation of a given unit. 
The process iterates, updating parameter estimates and sampling plans as data accumulate, until the design life is reached.

\subsection{Spatial Active Learning: Uniform Coverage via Space-Filling Design} 

Due to resource constraints, only $c$ out of $L$ units can be monitored at each of $o$ epochs. 
Let $\mathbf{W}$ be an $L \times o$ observation matrix, where $\mathbf{W}_{j,k} = 1$ indicates that unit $j$ is observed at epoch $k$, and $0$ otherwise. 
We aim to select the observed units such that the sampling elements of $\mathbf{W}$ satisfy space‑filling properties.

This problem can be transformed into selecting points in the normalized domain $[0, 1]^2$. 
The candidate set is defined as
\begin{equation*}
\mathcal{G} = \left\{ (u_k, v_j) : u_k = \frac{k - 0.5}{o},\; v_j = \frac{j - 0.5}{L},\;
k = 1,\dots, o,\; j = 1,\dots, L \right\},
\end{equation*}
where $u_k$ represents normalized time and $v_j$ normalized unit index. 
For any point $\boldsymbol{s} = (s_1, s_2) \in \mathcal{S}$, where $\mathcal{S} \subset \mathcal{G}$ is a subset, we set $\mathbf{W}_{j,k} = 1$ with $j = \lceil s_2 L \rceil$ and $k = \lceil s_1 o \rceil$. 
Our objective is to select $\mathcal{S}$ of cardinality $|\mathcal{S}| = oc$, subject to the constraint that each column of $\mathbf{W}$ sums to exactly $c$ (i.e., exactly $c$ units are observed at each epoch).

To evaluate design uniformity, we adopt the wrap‑around $L_2$ discrepancy (WD) \citep{fang2001wrap} as the optimality criterion. 
Compared to other discrepancy measures, WD is particularly suitable for this spatial sampling problem.
It possesses translation invariance, effectively treating the domain as a torus without physical boundaries, thereby eliminating boundary effects and ensuring that sampling points near the edges are weighted equally. 
Moreover, WD has a simple closed‑form expression that is computationally tractable.

For a design $\mathcal{S} = \{\boldsymbol{s}_1, \dots, \boldsymbol{s}_n\}$ with $n = oc$ and $\boldsymbol{s}_p = (s_{p,1}, s_{p,2})$, the squared WD is
\begin{equation*}
[\mathrm{WD}(\mathcal{S})]^2 = -\frac{16}{9} + \frac{1}{n^2}\sum_{p=1}^n \sum_{q=1}^n \prod_{d=1}^2 \phi(s_{p,d}, s_{q,d}),
\end{equation*}
where $\phi(x, y) = \frac{3}{2} - |x - y| + |x - y|^2$. 
Let $\mathcal{G}^\star$ denote the set of all valid subsets satisfying the column constraint. 
The optimal design for unit selection is found by
\begin{equation*}
\mathcal{S}^\star = \arg\min_{\mathcal{S} \subset \mathcal{G}^\star} [\mathrm{WD}(\mathcal{S})]^2.
\end{equation*}
Optimization algorithms such as threshold accepting or random swap are employed to search for $\mathcal{S}^\star$. 
Once $\mathcal{S}^\star$ is obtained, it yields the corresponding observation matrix $\mathbf{W}^\star$.

\begin{figure}
    \centering
    \includegraphics[width=\linewidth]{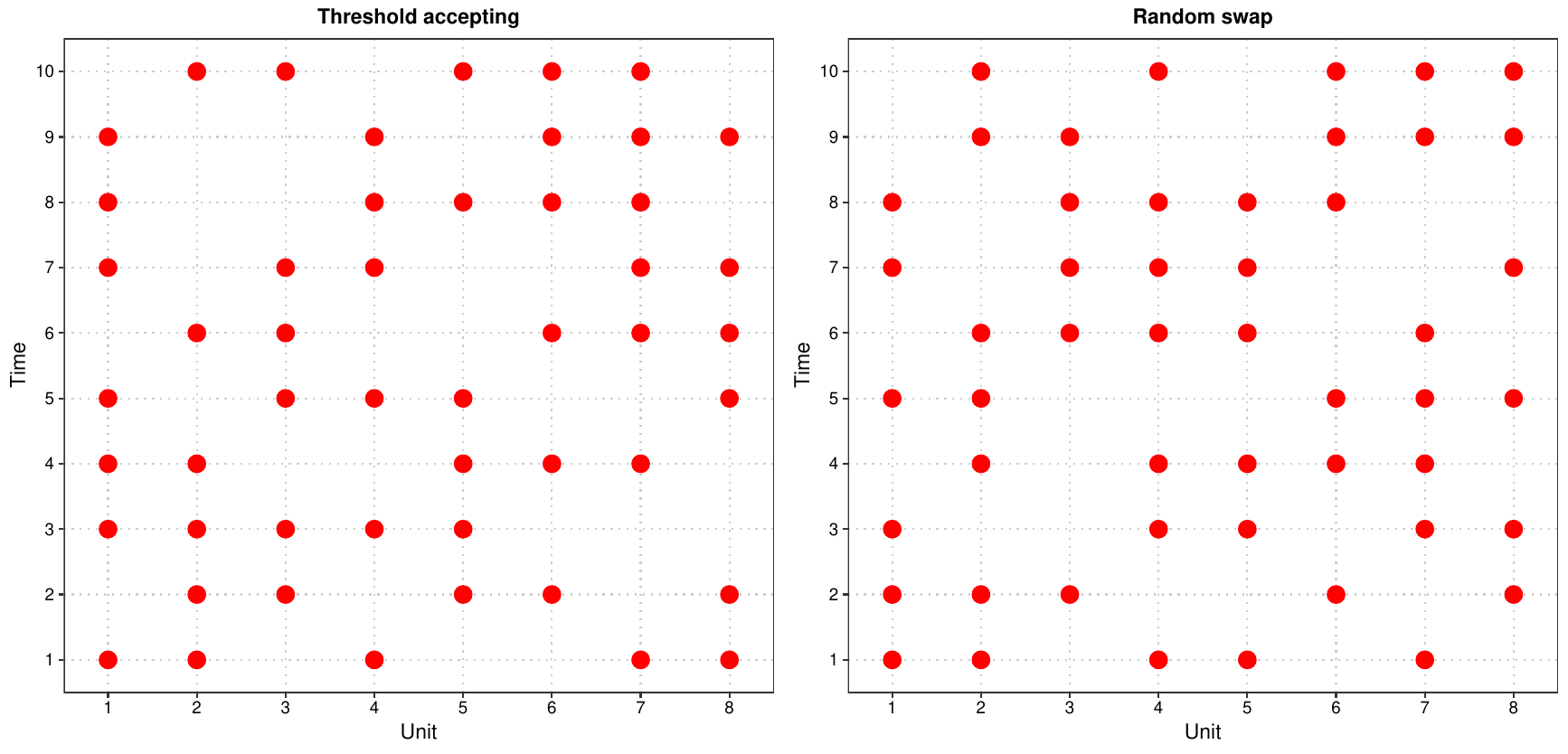}
    \caption{Illustration of the proposed spatial active learning via threshold accepting (left) and random swap algorithms (right) for a toy example with $L=8$, $o=10$, and $c=5$. Red points denote selected units.}
    \label{fig:toyunitsel}
\end{figure}

Figure~\ref{fig:toyunitsel} illustrates a toy example with $L = 8$, $o = 10$, and $c = 5$. 
The selected units are uniformly distributed throughout the domain. 
Vertically, the design strictly adheres to the resource constraint, with exactly $c = 5$ units selected at each epoch. 
Horizontally, the observation frequency is balanced among the $L = 8$ units, ensuring that no single unit is neglected or over‑sampled. 
The proposed criterion successfully avoids clustering and edge effects, thereby guaranteeing high‑quality data under limited resources.

\subsection{Sequential Temporal Active Learning: A Balanced Information Criterion}
During the early orbital phase, degradation trends are often subtle, so sampling is typically performed at regular intervals (e.g., every six months). 
After the initial monitoring phase, engineers may wish to optimize the time of the next observation $t_{m+1}$ given $m$ observations for a unit. 
For simplicity, we drop the subscript $i$ in this section, as the method is identical for all units.

Let $\boldsymbol{x}_\star = (\boldsymbol{x}^\top, X(t_{m+1}))^\top$ denote the augmented observation vector including the potential future observation, and let starred quantities denote those based on $X(t_{m+1})$. 
Accurate reliability prediction requires precise estimation of the key parameters $\boldsymbol{\zeta} = (\alpha, \gamma_1, \gamma_2)^\top$. 
This leads to a sequential D‑optimal design problem: choose $t_{m+1}$ to maximize the determinant of the Fisher information matrix (FIM) for $\boldsymbol{\zeta}$.

% The log‑likelihood for $\boldsymbol{x}_\star$ is
% \begin{equation*}
% \begin{aligned}
% l_\star(\boldsymbol{\theta} \mid \boldsymbol{x}_\star) \propto &-\ln\tau_a - \frac{1}{2}\ln|\widetilde{\boldsymbol{\Sigma}}_\star| - \frac{1}{2\tau_a^2} (\boldsymbol{x}_\star - \mu_a \boldsymbol{\Xi}_\star)^\top \widetilde{\boldsymbol{\Sigma}}_\star^{-1} (\boldsymbol{x}_\star - \mu_a \boldsymbol{\Xi}_\star),
% \end{aligned}
% \end{equation*}
% where $\widetilde{\boldsymbol{\Sigma}}_\star = \boldsymbol{\Xi}_\star \boldsymbol{\Xi}_\star^\top + \kappa^2 \boldsymbol{Q}_\star$ is the scaled covariance matrix for the augmented data. 
% To compute the FIM, we need the first‑order gradient of $l_\star$ with respect to $\boldsymbol{\zeta}$.

The log-likelihood function for the augmented vector is
\begin{align*}
    l_\star(\boldsymbol{\theta} \mid \boldsymbol{x}_\star) \propto &- \ln \tau_a - \frac{1}{2}\ln \vert \widetilde{\mathbf \Sigma}_\star \vert -\frac{1}{2\tau_a^2} \left(\boldsymbol{x}_\star - \mu_a \boldsymbol \Xi_\star\right)^\top\widetilde{\mathbf \Sigma}_\star^{-1}\left(\boldsymbol{x}_\star - \mu_a \boldsymbol \Xi_\star\right),
\end{align*}
with a $m_\star \times m_\star$ matrix $\widetilde{\mathbf{\Sigma}}_\star=\mathbf{\Sigma}_\star/\tau_a^2 = \boldsymbol\Xi_\star \boldsymbol\Xi_\star^{\top} + \kappa^2 \boldsymbol Q_\star$.
To compute the FIM, we require the first-order gradient of the log-likelihood with respect to $\boldsymbol{\zeta}$.

\subsubsection{Gradient with Respect to $\boldsymbol{\gamma}=(\gamma_1,\gamma_2)^\top$}
Since $\boldsymbol{\gamma}$ affects the likelihood only through $\boldsymbol{\Xi}_\star$, we first derive gradients with respect to elements of $\boldsymbol{\Xi}_\star$. 
Without loss of generality, we analyze the first element $\eta_1 = \lambda(t_1)\Lambda(t_1)$.
Gradients for the remaining elements are analogous. 
Using the decomposition from \citet{peng2009mis},
$$
|\widetilde{\boldsymbol{\Sigma}}_\star| = |\kappa^2 \boldsymbol{Q}_\star| \bigl(1 + \boldsymbol{\Xi}_\star^\top (\kappa^2 \boldsymbol{Q}_\star)^{-1} \boldsymbol{\Xi}_\star\bigr),
$$
and
$$
\widetilde{\boldsymbol{\Sigma}}_\star^{-1} = \widetilde{\boldsymbol{Q}}_\star^{-1} - \frac{\widetilde{\boldsymbol{Q}}_\star^{-1} \boldsymbol{\Xi}_\star \boldsymbol{\Xi}_\star^\top \widetilde{\boldsymbol{Q}}_\star^{-1}}{1 + \boldsymbol{\Xi}_\star^\top \widetilde{\boldsymbol{Q}}_\star^{-1} \boldsymbol{\Xi}_\star},
$$
where $\widetilde{\boldsymbol{Q}}_\star = \kappa^2 \boldsymbol{Q}_\star$. Define $B_\star = 1 + \boldsymbol{\Xi}_\star^\top \widetilde{\boldsymbol{Q}}_\star^{-1} \boldsymbol{\Xi}_\star$ and $\mathbf{A}_\star = \widetilde{\boldsymbol{Q}}_\star^{-1} \boldsymbol{\Xi}_\star \boldsymbol{\Xi}_\star^\top \widetilde{\boldsymbol{Q}}_\star^{-1}$. The required derivatives are then
\begin{align*}
\frac{\partial \ln|\widetilde{\boldsymbol{\Sigma}}_\star|}{\partial \eta_1} &= \frac{2 \boldsymbol{e}_1^\top \widetilde{\boldsymbol{Q}}_\star^{-1} \boldsymbol{\Xi}_\star}{B_\star}, \\
\frac{\partial \widetilde{\boldsymbol{\Sigma}}_\star^{-1}}{\partial \eta_1} &=
\frac{2(\boldsymbol{e}_1^\top \widetilde{\boldsymbol{Q}}_\star^{-1} \boldsymbol{\Xi}_\star) \mathbf{A}_\star}{B_\star^2}
- \frac{\widetilde{\boldsymbol{Q}}_\star^{-1} (\boldsymbol{e}_1 \boldsymbol{\Xi}_\star^\top + \boldsymbol{\Xi}_\star \boldsymbol{e}_1^\top) \widetilde{\boldsymbol{Q}}_\star^{-1}}{B_\star}, \\
\frac{\partial D_\star}{\partial \eta_1} &= -2\mu_a \boldsymbol{e}_1^\top \widetilde{\boldsymbol{\Sigma}}_\star^{-1} (\boldsymbol{x}_\star - \mu_a \boldsymbol{\Xi}_\star) \nonumber + (\boldsymbol{x}_\star - \mu_a \boldsymbol{\Xi}_\star)^\top \frac{\partial \widetilde{\boldsymbol{\Sigma}}_\star^{-1}}{\partial \eta_1} (\boldsymbol{x}_\star - \mu_a \boldsymbol{\Xi}_\star),
\end{align*}
where $\boldsymbol{e}_1$ is a column vector with the first element being 1 and the remaining elements being 0, $D_\star $ is the quadratic form in the likelihood. 
Combining these,
$$
\frac{\partial l_\star}{\partial \eta_1} = -\frac{1}{2}\frac{\partial \ln|\widetilde{\boldsymbol{\Sigma}}_\star|}{\partial \eta_1} - \frac{1}{2\tau_a^2}\frac{\partial D_\star}{\partial \eta_1}.
$$
The chain rule then gives $\partial l_\star/\partial \boldsymbol{\gamma}$ via
$$
\frac{\partial \boldsymbol{\Xi}_\star}{\partial \boldsymbol{\gamma}^\top} =
\operatorname{diag}\bigl(\lambda(t_1)\Lambda(t_1), \dots, \lambda(t_{m+1})\Lambda(t_{m+1})\bigr)
\begin{pmatrix}
z_1(t_1) & z_2(t_1)\\
\vdots & \vdots\\
z_1(t_{m+1}) & z_2(t_{m+1})
\end{pmatrix}.
$$

\subsubsection{Gradient with Respect to $\alpha$}
For $\alpha$, we again use the decomposition $\ln|\widetilde{\boldsymbol{\Sigma}}_\star| = \ln|\widetilde{\boldsymbol{Q}}_\star| + \ln B_\star$ and $\widetilde{\boldsymbol{\Sigma}}_\star^{-1} = \widetilde{\boldsymbol{Q}}_\star^{-1} - \mathbf{A}_\star / B_\star$. 
The determinant of $\widetilde{\boldsymbol{Q}}_\star$ has the closed form
$$
|\widetilde{\boldsymbol{Q}}_\star| = \kappa^{2(m+1)} \prod_{k=1}^{m+1} (t_k^\alpha - t_{k-1}^\alpha), \quad t_0 = 0.
$$
Thus, 
$$
\frac{\partial |\widetilde{\boldsymbol{Q}}_\star|}{\partial \alpha} = |\widetilde{\boldsymbol{Q}}_\star|
\Biggl( \ln t_1 + \sum_{k=2}^{m+1} \frac{t_k^\alpha \ln t_k - t_{k-1}^\alpha \ln t_{k-1}}{t_k^\alpha - t_{k-1}^\alpha} \Biggr).
$$

The inverse $\widetilde{\boldsymbol{Q}}_\star^{-1}$ is tridiagonal with entries expressed in terms of $d_k = t_k^\alpha - t_{k-1}^\alpha$. 
Its derivative $\partial \widetilde{\boldsymbol{Q}}_\star^{-1} / \partial \alpha$ is obtained by differentiating each entry, using $d_k' = t_k^\alpha \ln t_k - t_{k-1}^\alpha \ln t_{k-1}$. 
The derivatives of $B_\star$, $\mathbf{A}_\star$, and ultimately $\widetilde{\boldsymbol{\Sigma}}_\star^{-1}$ follow from these building blocks. 
In particular,
$$
\frac{\partial B_\star}{\partial \alpha} = 2\frac{\partial \boldsymbol{\Xi}_\star^\top}{\partial \alpha} \widetilde{\boldsymbol{Q}}_\star^{-1} \boldsymbol{\Xi}_\star + \boldsymbol{\Xi}_\star^\top \frac{\partial \widetilde{\boldsymbol{Q}}_\star^{-1}}{\partial \alpha} \boldsymbol{\Xi}_\star,
$$
and
$$
\frac{\partial D_\star}{\partial \alpha} = -2\mu_a \frac{\partial \boldsymbol{\Xi}_\star^\top}{\partial \alpha} \widetilde{\boldsymbol{\Sigma}}_\star^{-1} (\boldsymbol{x}_\star - \mu_a \boldsymbol{\Xi}_\star) + (\boldsymbol{x}_\star - \mu_a \boldsymbol{\Xi}_\star)^\top \frac{\partial \widetilde{\boldsymbol{\Sigma}}_\star^{-1}}{\partial \alpha} (\boldsymbol{x}_\star - \mu_a \boldsymbol{\Xi}_\star).
$$
Finally,
$$
\frac{\partial l_\star}{\partial \alpha} = -\frac{1}{2}\frac{\partial \ln|\widetilde{\boldsymbol{\Sigma}}_\star|}{\partial \alpha} - \frac{1}{2\tau_a^2}\frac{\partial D_\star}{\partial \alpha}.
$$

\subsubsection{The Proposed Criterion}

Under standard regularity conditions, the FIM $\mathcal{I}(\boldsymbol{\zeta} \mid \boldsymbol{x}_\star)$ for $\boldsymbol{\zeta}=(\alpha,\gamma_1,\gamma_2)^\top$ can be derived. 
Its $(l,k)$th element is
$$
\mathcal{I}_{l,k} = \int \frac{\partial l_\star(\widehat{\boldsymbol{\theta}} \mid \boldsymbol{x}_\star)}{\partial \zeta_l} \frac{\partial l_\star(\widehat{\boldsymbol{\theta}} \mid \boldsymbol{x}_\star)}{\partial \zeta_k} \, f\bigl(X(t_{m+1}) \mid \boldsymbol{x}\bigr) \, dX(t_{m+1}).
$$

The degradation path follows a power law $\Lambda(t) = t^\alpha$. 
For $\alpha > 1$, the trajectory exhibits a slow initial phase followed by rapid acceleration. 
A pure D‑optimal design, which maximizes $\det \mathcal{I}$, would favor sampling at the boundaries, i.e., the end‑of‑life phase, where the signal is strongest. 
This risks clustering all observations in the final stages and missing the onset of acceleration. 
To obtain robust estimates of the curvature parameter $\alpha$, we introduce a penalty term that encourages exploration of the transition phase. 
Let $\Lambda'(t) = \alpha t^{\alpha-1}$ be the instantaneous degradation rate. 
The term $1/\Lambda'(t)$ is largest early in life, providing an exploration bonus for interior points. The proposed comprehensive criterion is
\begin{equation}
\label{eq:twostage_cri}
\Gamma(t_{m+1}) = \omega_1 \left| \frac{\ln \det \mathcal{I}(\boldsymbol{\zeta} \mid \boldsymbol{x}_\star)}{\ln \det \mathcal{I}(\widehat{\boldsymbol{\zeta}} \mid \boldsymbol{x}_\star)} \right| + \frac{\omega_2}{\Lambda'(t_{m+1} \mid \widehat{\alpha})},
\end{equation}
where $\widehat{\boldsymbol{\zeta}}$ denotes current estimates, and $\omega_1, \omega_2 \ge 0$, $\omega_1 + \omega_2 = 1$, balance information gain versus exploration. 
For $0 < \alpha < 1$, the second term is replaced by $\omega_2 \Lambda'(t_{m+1} \mid \widehat{\alpha})$ to favor sampling during periods of increasing degradation rate. 
The optimal future sampling time is
$$
t_{m+1}^\star = \arg\max_{t_{m+1} > t_m} \Gamma(t_{m+1}).
$$

\section{Illustrative Examples}
\label{sec:example}
\subsection{Artificial Example}
In this subsection, we conduct a series of artificial experiments to evaluate the performance of the proposed degradation modeling framework based on spatiotemporal active learning.
The simulation procedure mimics the lifecycle of electronic units of satellites under the varying working environment.
Synthetic degradation paths are generated based on the proposed model~\eqref{eq:base_model} with the true parameters.
The simulation considers $L=5$ units with $c=3$ units to observe every time. 
The true model parameters are set to $\mu_a=1$, $\tau_a=0.1$, $\kappa=1$, $\rho=0.5$, $\gamma_1=0.1$, and $\gamma_2=0.2$, leading to the designed lifespans of 10 years.
The iterative simulation process is structured as follows.

First, initial data from the early orbital phase are generated.
Specifically, since the reliability of such electronic devices operating over long periods remains stable during the initial years \citep{wang2022adaptive}, frequent monitoring of their degradation is not necessary.
Combined with engineering experience, we generate one observation every half a year during the first 5 years.
Based on this dataset, the model parameters are estimated, which subsequently drive the active learning strategy.
Given the optimized subset of units and the next sampling time, new observations are generated from the conditional multivariate normal distribution, preserving the spatiotemporal dependence structure.
This cycle of estimation, active learning, and data generation is iterated until the design life (namely, 10 years) is reached.
Finally, the reliability over 11-12 years of each unit is assessed using the Monte Carlo method, which is of intrinsic interest to the engineers.
The whole simulation procedure is replicated 1000 times.

To ensure a thorough evaluation, we employ four experimental settings as follows.
\begin{enumerate}
    \item \textbf{Degradation path shape ($S_1$):} $\alpha \in \{0.5, 1.2\}$, representing concave ($S_1=0$) and convex ($S_1=1$) degradation trends, respectively.
    % Corresponding critical thresholds for reliability analysis are 12 with $\alpha=0.5$ and 65 with $\alpha=1.2$. 
    \item \textbf{Initial unit selection ($S_2$):} Whether the proposed unit selection is applied during the initial data collection phase, e.g., the first 5 years. ($S_2=1$ denotes implementation; $S_2=0$ denotes otherwise).
    \item \textbf{Sampling plan ($S_3$):} Whether the sampling time is unconstrained ($S_3=0$) or restricted ($S_3=1$) to a selection scheme that contains consecutive non-overlapping intervals with the length $\Delta t$ that are predetermined by engineering experience.
    In this study, engineering experience suggests that observations are scheduled semi-annually from years 5 to 8 ($\Delta t=0.5$), and quarterly from years 8 to 10 ($\Delta t=0.25$), resulting in a total of 14 time points.
    \item \textbf{Parameter uncertainty ($S_4$):} Whether the shape parameter $\alpha$ is treated as known ($S_4=1$) or not ($S_4=0$).
\end{enumerate}
For detailed denotations for all scenario configurations, please refer to the Appendix.

We denote the proposed model via our spatiotemporal active learning strategy as $M_0$.
Given an existing observation plan, $M_0$ selects the most suitable sampling time within each prescribed interval ($S_3=0$); otherwise, it determines sampling times for the years 5-10.
Two methods are further applied as competitors.
The first one incorporates the proposed model with a traditional uniform temporal sampling strategy, denoted by $M_1$.
Where a predetermined selection scheme exists, it selects the right endpoint of each given interval, according to traditional engineering sampling methods. 
For an unconstrained sampling plan, the number of observations is kept similar to $M_0$ to ensure a fair comparison between $M_0$ and $M_1$.
This strategy serves as a comparative baseline to evaluate the efficiency of the proposed temporal active learning framework, especially when there is no prior knowledge about the sampling.
The other one is the traditional method, denoted as $M_2$.
It assumes independence among units, monitors all units ($c=L$), and employs temporal sampling via the traditional engineering sampling plan (e.g., every half a year over 5-10 years).

First, we conduct a quantitative comparison between the proposed spatiotemporal active learning strategy ($M_0$) and the uniform strategy ($M_1$), which are both based on our model, to demonstrate the efficiency of the temporal active learning criterion.
% We select three representative scenarios, $C_{1}$, $C_{14}$, and $C_{11}$, for detailed analysis.
The performance is evaluated by calculating the mean relative error between the predicted reliability and the true values, which is computed across all 5 units.
% Corresponding average relative errors over 1000 replications are summarized in Table~\ref{tab:resultscom}.

We begin by analyzing two scenarios where $(S_1,S_2,S_3,S_4)$ is $(1,0,1,0)$ or $(0,0,1,0)$.
Specifically, observation intervals are predetermined, and the proposed unit selection is applied only during 5-10 years, with $\alpha=1.2$ or $0.5$.
These scenarios are designed to evaluate how the methods perform under the actual engineering strategies employed.
As shown in Table~\ref{tab:resultscom1}, even under these strict constraints with different values of $\alpha$, $M_0$ demonstrates superior performance.
For instance, $M_0$ achieves a remarkably low initial relative error of 0.057\%, compared to 0.060\% for $M_1$.
This indicates that even when sampling windows are fixed (e.g, two methods use the same number of observations), optimizing the specific time within those intervals based on model information yields benefits over uniformly sampling, which is commonly used in real cases.

\begin{table}[!htbp]
\centering
  \caption{Average relative errors over 1000 replications ($\%$) with $(S_1,S_2,S_3,S_4)=$ $(1,0,1,0)$ and $(0,0,1,0)$.}
  \label{tab:resultscom1}
\begin{tabular}{ccc|cc}
\hline\hline
& \multicolumn{2}{c|}{$(1,0,1,0)$} &  \multicolumn{2}{c}{$(0,0,1,0)$} \\
   Time    & $M_0$                 & $M_1$      & $M_0$                 & $M_1$      \\
       \hline\
10.125 & \textbf{0.057031} & 0.060101 & \textbf{0.831476} & 0.990044 \\
10.25  & \textbf{0.106216} & 0.116602 & \textbf{1.160813} & 1.351985 \\
10.375 & \textbf{0.193282} & 0.217947 & \textbf{1.604081} & 1.841957 \\
10.5   & \textbf{0.32953}  & 0.378337 & \textbf{2.131811} & 2.424643 \\
10.625 & \textbf{0.881492} & 1.034227 & \textbf{2.825481} & 3.174507 \\
10.75  & \textbf{1.374355} & 1.605227 & \textbf{3.653572} & 4.034231 \\
10.875 & \textbf{2.059839} & 2.395623 & \textbf{4.681642} & 5.110656 \\
11     & \textbf{2.947713} & 3.415247 & \textbf{5.754727} & 6.229326 \\
11.125 & \textbf{5.428949} & 6.210558 & \textbf{6.984299} & 7.514318 \\
11.25  & \textbf{6.980237} & 7.950766 & \textbf{8.297449} & 8.841914 \\
11.375 & \textbf{8.869664} & 9.980824 & \textbf{9.702549} & 10.28248 \\
11.5   & \textbf{10.89951} & 12.19788 & \textbf{11.02037} & 11.61823 \\
11.625 & \textbf{15.41957} & 17.03213 & \textbf{12.40448} & 13.05386 \\
11.75  & \textbf{18.02554} & 19.79833 & \textbf{13.89518} & 14.51982 \\
11.875 & \textbf{20.65199} & 22.58804 & \textbf{15.19242} & 15.85698 \\
12     & \textbf{23.32317} & 25.38593 & \textbf{16.48186} & 17.13989 \\
\hline\hline
\end{tabular}
\end{table}

Subsequently, for high-reliability electronic devices, in fact, a more effective approach does not require observing all units during the initial stage.
Therefore, we consider scenarios with predetermined observation intervals and apply the proposed unit selection strategy over the entire 0‑10‑year lifecycle, specifically with $(S_1,S_2,S_3,S_4) = (1,1,1,1)$ or $(0,1,1,1)$.
Table \ref{tab:resultscom2} summarizes the average relative errors of $M_0$ and $M_1$ across 1000 replications. 
The results demonstrate that $M_0$ outperforms $M_1$ throughout most of the degradation process. 
For example, for $10.75$ years, $M_0$ achieves a relative error of 1.508\%, whereas $M_1$ yields 1.690\%.
Notably, during the final stage of the lifecycle, the performance of the two models becomes comparable. 
For $11.75$ years, the average relative error of $M_1$ (22.058\%) is slightly higher than the value of $M_0$ (21.930\%), and this trend continues for $12.0$ years, where $M_1$ records 29.821\% against $M_0$, which yields 30.006\%. 

\begin{table}[!htbp]
\centering
  \caption{Average relative errors over 1000 replications ($\%$) with $(S_1,S_2,S_3,S_4) = (1,1,1,1)$ and $(1,0,1,1)$.}
  \label{tab:resultscom2}
\begin{tabular}{ccc|cc}
\hline\hline
    & \multicolumn{2}{c|}{$(1,1,1,1)$} &  \multicolumn{2}{c}{$(1,0,1,1)$} \\
   Time    & $M_0$                 & $M_1$      & $M_0$                 & $M_1$      \\
       \hline
10.125 & \textbf{0.054293} & 0.067368 & \textbf{0.545085} & 0.630219 \\
10.250 & \textbf{0.121192} & 0.146095 & \textbf{0.823907} & 0.923724 \\
10.375 & \textbf{0.241093} & 0.286382 & \textbf{1.162903} & 1.280622 \\
10.500 & \textbf{0.472905} & 0.548444 & \textbf{1.606356} & 1.741849 \\
10.625 & \textbf{0.881574} & 1.005022 & \textbf{2.183721} & 2.336306 \\
10.750 & \textbf{1.508154} & 1.689699 & \textbf{2.972111} & 3.136299 \\
10.875 & \textbf{2.562045} & 2.827282 & \textbf{3.973253} & 4.128645 \\
11.000 & \textbf{3.935440} & 4.273591 & \textbf{5.071165} & 5.210964 \\
11.125 & \textbf{5.912405} & 6.299959 & \textbf{6.317419} & 6.432001 \\
11.250 & \textbf{8.329783} & 8.733579 & \textbf{7.765005} & 7.865554 \\
11.375 & \textbf{11.241422} & 11.634243 & \textbf{9.281586} & 9.343763 \\
11.500 & \textbf{14.429876} & 14.786094 & \textbf{10.772534} & 10.804214 \\
11.625 & \textbf{17.998704} & 18.243628 & \textbf{12.344846} & 12.346398 \\
11.750 & \textbf{21.930173} & 22.058294 & 14.129067 & \textbf{14.080957} \\
11.875 & 26.055526 & \textbf{26.033486} & 15.649668 & \textbf{15.579246} \\
12.000 & 30.006021 & \textbf{29.821432} & 17.188731 & \textbf{17.115164} \\
\hline\hline
\end{tabular}
\end{table}

Then, we examine the scenario in which observation intervals are not predetermined, and the value of $\alpha$ is unknown, applying the proposed spatial active learning strategy throughout the full lifecycle (0-10 years).
This scenario is intended to emulate real-world conditions, particularly those in which data transmission is limited, and engineers are uncertain about the appropriate selection scheme.
Consequently, either non‑informative approaches (e.g., uniform sampling) or sampling based on model uncertainty and the rate of degradation (i.e., the derivative of the degradation function) can be employed.
$(S_1,S_2,S_3,S_4)$ is $(1,1,0,0)$ and $(0,1,0,0)$, respectively.
Despite these complexities, $M_0$ maintains superior accuracy, reducing the final-stage error from 30.108\% ($M_1$) to 29.132\% ($M_1$).
% Crucially, it is worth noting that this performance improvement is achieved with significantly reduced data consumption.
% In this scenario, $M_0$ utilizes approximately 55 observations on average, whereas $M_1$ requires a fixed 60 observations.
Results under other scenarios shown in the Appendix also present that $M_0$ is comparable or even better than $M_1$.
This result highlights the core advantage of the proposed method: by leveraging the dynamic rule of degradation and model accuracy to maximize information gain, it can achieve higher prediction accuracy with no prior knowledge about the sample plan, while reducing the data size by selecting representative units.

\begin{table}[!htbp]
\centering
  \caption{Average relative errors over 1000 replications ($\%$) with $(S_1,S_2,S_3,S_4) = (1,1,0,0)$ and $(0,1,0,0)$.}
  \label{tab:resultscom3}
\begin{tabular}{ccc|cc}
\hline\hline
    & \multicolumn{2}{c|}{$(1,1,0,0)$} &  \multicolumn{2}{c}{$(0,1,0,0)$} \\
   Time    & $M_0$                 & $M_1$      & $M_0$                 & $M_1$      \\
       \hline
10.125 & \textbf{0.043734} & 0.058246 & \textbf{0.577743} & 0.662222 \\
10.250 & \textbf{0.104091} & 0.126442 & \textbf{0.861629} & 0.969886 \\
10.375 & \textbf{0.213826} & 0.247743 & \textbf{1.208568} & 1.348249 \\
10.500 & \textbf{0.436478} & 0.482058 & \textbf{1.660435} & 1.833644 \\
10.625 & \textbf{0.829029} & 0.899187 & \textbf{2.248105} & 2.462118 \\
10.750 & \textbf{1.439813} & 1.530540 & \textbf{3.041877} & 3.297767 \\
10.875 & \textbf{2.473795} & 2.583003 & \textbf{4.051456} & 4.356582 \\
11.000 & \textbf{3.807664} & 3.951631 & \textbf{5.139391} & 5.496345 \\
11.125 & \textbf{5.686434} & 5.918260 & \textbf{6.376490} & 6.793698 \\
11.250 & \textbf{8.022159} & 8.332700 & \textbf{7.825340} & 8.296817 \\
11.375 & \textbf{10.850785} & 11.284752 & \textbf{9.362042} & 9.883849 \\
11.500 & \textbf{13.970714} & 14.540027 & \textbf{10.861963} & 11.402644 \\
11.625 & \textbf{17.400912} & 18.160114 & \textbf{12.441025} & 13.025153 \\
11.750 & \textbf{21.250674} & 22.103612 & \textbf{14.214697} & 14.838724 \\
11.875 & \textbf{25.236935} & 26.246048 & \textbf{15.773258} & 16.405416 \\
12.000 & \textbf{29.132435} & 30.107902 & \textbf{17.343956} & 18.004287 \\
\hline\hline
\end{tabular}
\end{table}

\begin{figure}
    \centering
    \includegraphics[width=\linewidth]{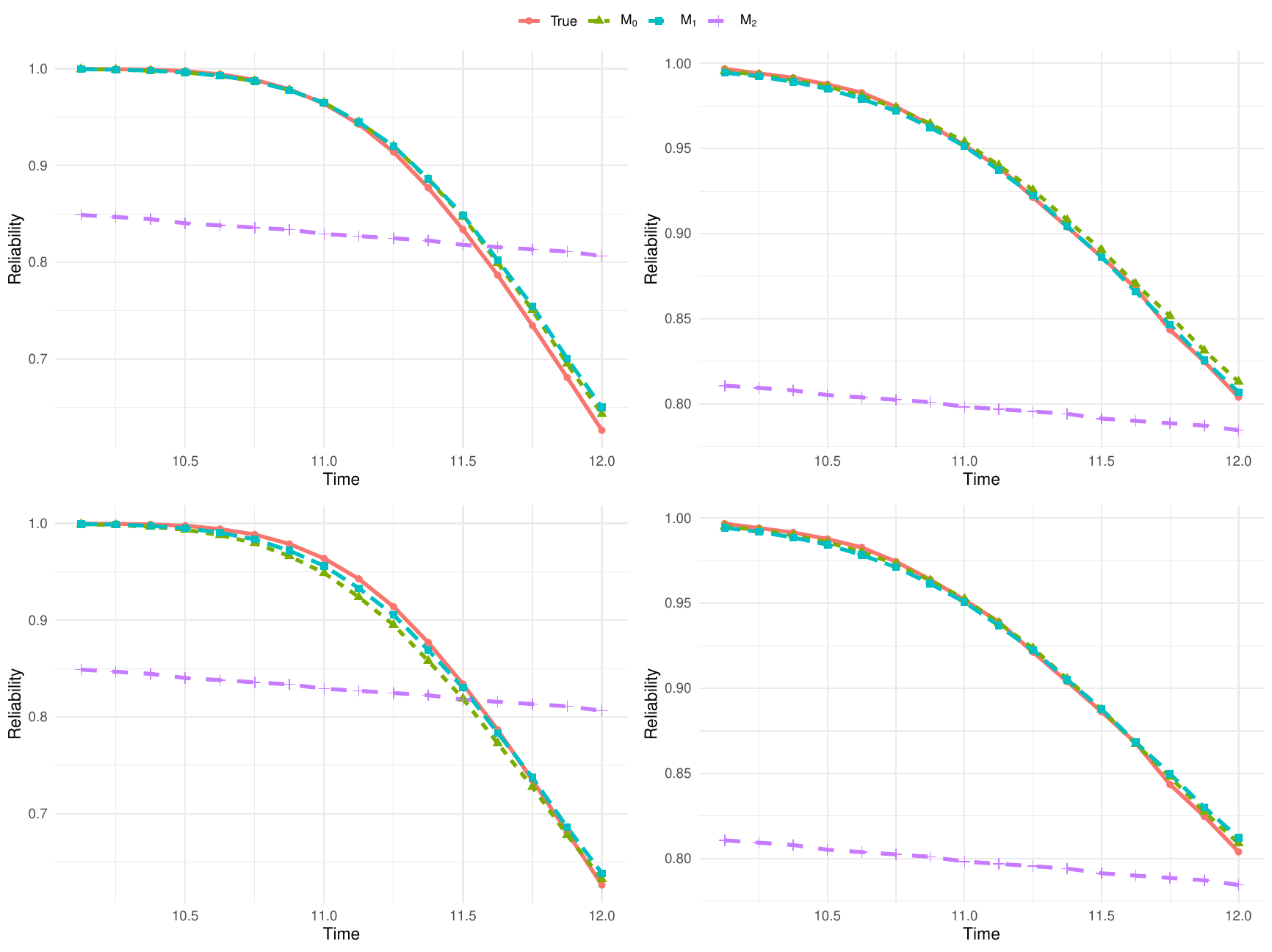}
    \caption{Average reliability prediction results over 1000 replications when $(S_1,S_2,S_3,S_4)$ is $(1,1,1,0)$, $(0,1,1,0)$, $(1,1,0,0)$, and $(0,1,0,0)$.}
    \label{fig:simcomex}
\end{figure}

In addition, we evaluate the impact of spatial dependence and assess the validity of the proposed degradation model.
We compare models $M_0$ and $M_1$, both of which adopt our degradation model, against $M_2$, which assumes unit independence and strictly adheres to the preset observation schedule.
Specifically, $M_2$ observes all units at each time point, with semi-annual sampling ($\Delta t=0.5$) from year 0 to 8 and quarterly sampling ($\Delta t=0.25$) from year 8 to 10, yielding a total of 24 observations.
% Four scenarios are considered by the degradation shape parameter $\alpha$ and whether its value is known or unknown.
To quantify the efficiency of the proposed model, we analyze four scenarios corresponding to $(S_1,S_2,S_3,S_4)$ set as $(1,1,1,0)$, $(0,1,1,0)$, $(1,1,0,0)$, and $(0,1,0,0)$, respectively.
These scenarios account for two distinct degradation curve shapes and whether the sampling plan is predetermined, with unknown $\alpha$.
To further demonstrate the novelty of our approach, we apply unit selection across the entire observation period when implementing $M_0$ and $M_1$, resulting in fewer required samples than $M_2$.
The average reliability is computed across 5 units per replication, and the mean over 1000 such replications, alongside the true reliability values, is shown in Figure~\ref{fig:simcomex}.
As seen, $M_2$, despite utilizing data from all units at all planned epochs (namely 120 observations), yields suboptimal reliability predictions.
Conversely, both the proposed method $M_0$ and the proposed model with the traditional uniform sampling strategy $M_1$ provide trajectories that closely track the true reliability.
This comparison underscores that reliability predictions solely on independent units (as in $M_2$) are insufficient and can lead to misleading prognostic results.
It is also worth noting that, while maintaining prediction accuracy, the proposed active learning strategy reduces the overall number of required observations to about 52, thereby significantly lowering data acquisition costs.

\subsection{Real Case Study}
This subsection presents a practical application of MOSFET devices, which are critical switching components in satellite PDUs. 
To validate the proposed framework under realistic on-orbit conditions, an emulator was developed by Beijing Spacecrafts based on extensive historical data \citep{wang2022adaptive}.
The system comprises $L=12$ spatially correlated units.
The model parameters are set as $\mu_a=4.661$, $\tau_a=0.405$, $\kappa=1.841$ (corresponding to $\sigma=0.735$), $\rho=0.5$, $\gamma_1=0.1$, and $\gamma_2=0.2$.
According to the working conditions involved, the corresponding observing times are selected until the next observing time is larger than 10 years.
The reliability prediction focuses on the 10-12 years.

To reflect the constraints of satellite data transmission, the maximum allowable observations per epoch are restricted to $c=4$ during the initial phase (0-5 years) and $c=6$ during the later phase (5-10 years).
Data is sampled semi-annually during the first 5 years.
Similar to the artificial examples, $M_2$ monitors all $L=12$ units according to a fixed schedule (semi-annually from years 5-8, quarterly from years 8-10) and assumes statistical independence among units.
Both $M_0$ and $M_1$ incorporate the proposed model and select a representative unit subset over the full lifecycle, with semi-annual sampling from years 0 to 5. 
The key distinction is that $M_0$ adaptively identifies optimal sampling times, while $M_1$ collects samples annually from years 5 to 10, achieving the same number of observations as $M_0$ for a fair comparison.

\begin{figure}
    \centering
    \includegraphics[width=\linewidth]{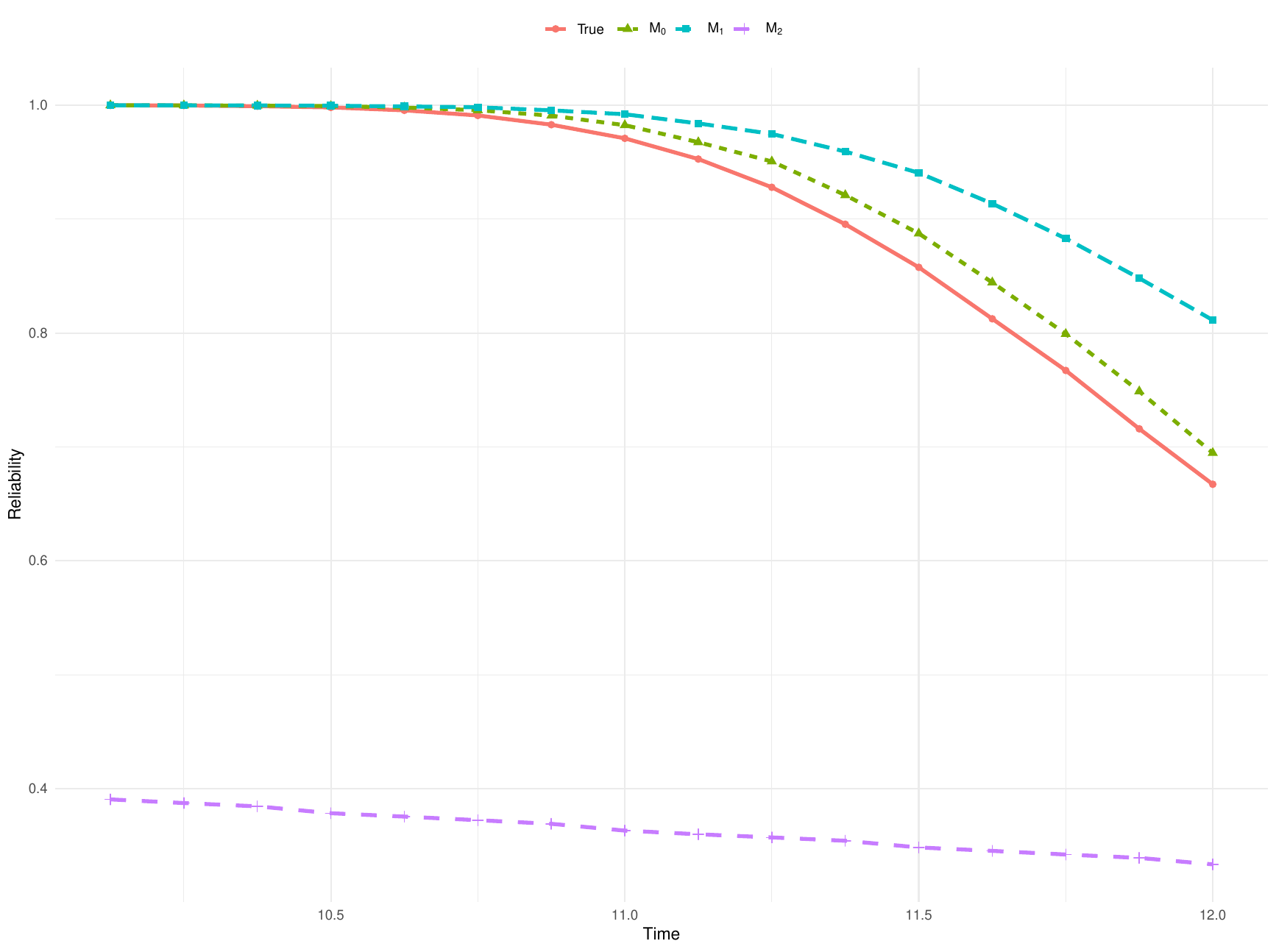}
    \caption{Average reliability predictions over 12 devices during 10-12 years.}
    \label{fig:rescase}
\end{figure}

The reliability prediction results are illustrated in Fig.~\ref{fig:rescase}, revealing a significant difference in predictive performance.
$M_2$, despite utilizing data from all 12 units and a total of 190 samples, yields a substantial underestimation of system reliability (predicting about $0.4$ versus the true value of around $1.0$ for the year 10.5).
This bias arises because $M_2$ neglects the spatial thermal coupling; by treating correlated units as independent, it fails to account for the synchronized degradation trends induced by the location topology, leading to overly conservative estimates.
In contrast, $M_0$ and $M_1$, each utilizing 70 samples, demonstrate superior predictive fidelity, producing reliability trajectories that closely align with the true values. 
Particularly, $M_0$ outperforms $M_1$, as it effectively leverages the incorporated model information.
% It suggests that our proposed framework $M_0$ can substantially reduce data acquisition costs with comparable accuracy.
This result confirms that the proposed model and spatiotemporal active learning strategy effectively capture the system's health status with minimal data consumption, offering a robust solution for resource-constrained satellite monitoring.

\section{Conclusion}
\label{sec:con}
This article has addressed the fundamental challenge of online reliability prediction for satellite electronics under severe data constraints, dynamic environments, and spatial dependence, advancing the state of the art through three integrated contributions: a hierarchical Wiener process model that simultaneously captures time-varying environmental covariates, unit-to-unit heterogeneity, and spatial correlation among adjacent units via correlated random coefficients; an efficient profile likelihood estimation procedure that concentrates out scale parameters analytically to reduce computational dimensionality; and most critically, a two-stage active learning strategy that optimizes both spatial selection—using wrap-around $L_2$ discrepancy to ensure uniform coverage across units—and temporal scheduling—via a balanced information criterion that explicitly trades off D‑optimal information gain against exploration of the critical degradation transition phase, overcoming a fundamental limitation of pure boundary-favoring designs.
A series of artificial examples and the application to the PDUs demonstrate the efficiency and practical utility of this framework, making it a promising solution for the prognostic and health management of high-value, long-endurance aerospace electronic equipment.

Future research directions may include extending this framework to more complex spatial layouts, multiple dependent performance characteristics with competing failure modes, or exploring a more efficient model, such as physics-informed neural networks, to further enhance degradation trend capture.

\section*{Disclosure statement}
\label{disclosure-statement}
No potential conflict of interest was reported by the authors.

% \section{Data Availability Statement}
% \label{data-availability-statement}

% Deidentified data have been made available at the following URL: XX.

% \phantomsection\label{supplementary-material}
% \bigskip

% \begin{center}

% {\large\bf SUPPLEMENTARY MATERIAL}

% \end{center}

% \begin{description}
% \item[Title:]
% Brief description. (file type)
% \item[R-package for MYNEW routine:]
% R-package MYNEW containing code to perform the diagnostic methods
% described in the article. The package also contains all datasets used as
% examples in the article. (GNU zipped tar file)
% \item[HIV data set:]
% Data set used in the illustration of MYNEW method in
% Section~\ref{sec-verify} (.txt file).
% \end{description}

\bibliography{ref.bib}

\end{document}